# Galactic diffuse gamma rays — recalculation based on the new measurements of cosmic electron spectrum

Juan Zhang[1], Qiang Yuan[1], Xiao-Jun Bi[1,2]

[1] *Key Laboratory of Particle Astrophysics, Institute of High Energy Physics, Chinese Academy of Sciences, Beijing 100049, P. R. China*
[2] *Center for High Energy Physics, Peking University, Beijing 100871, P. R. China*

## ABSTRACT

In this work, we revisit the all-sky Galactic diffuse $\gamma$-ray emission taking into account the new measurements of cosmic ray electron/positron spectrum by PAMELA, ATIC and Fermi, which show excesses of cosmic electrons/positrons beyond the expected fluxes in the conventional model. Since the origins of the extra electrons/positrons are not clear, we consider three different scenarios to account for the excesses: the astrophysical sources such as the Galactic pulsars, dark matter decay and annihilation. Further, new results from Fermi-LAT of the (extra-)Galactic diffuse $\gamma$-ray are adopted.

The background cosmic rays without the new sources give lower diffuse $\gamma$ rays compared to Fermi-LAT observation, which is consistent with previous analysis. The scenario with astrophysical sources predicts diffuse $\gamma$-rays with little difference with the background. The dark matter annihilation models with $\tau^{\pm}$ final state are disfavored by the Fermi diffuse $\gamma$-ray data, while there are only few constraints on the decaying dark matter scenario. Furthermore, these is always a bump at higher energies ($\sim$ TeV) of the diffuse $\gamma$-ray spectra for the dark matter scenarios due to final state radiation. Finally we find that the Fermi-LAT diffuse $\gamma$-ray data can be explained by simply enlarging the normalization of the electron spectrum without introduce any new sources, which may indicate that the current constraints on the dark matter models can be much stronger given a precise background estimate.

*Subject headings:* cosmic rays – dark matter – gamma rays:diffuse background

## 1. Introduction

The diffuse Galactic $\gamma$-ray is an important issue in astrophysics, since it provides a direct measurement of cosmic rays (CRs) at distant locations. It provides a crucial diagnostic of the transportation process of CRs in the interstellar space. Further it forms the foreground of the extragalactic diffuse $\gamma$-ray emission. The diffuse Galactic $\gamma$-rays may also provide information of exotic sources, such as the dark matter (DM) annihilation or decay.

The diffuse Galactic $\gamma$-rays are generally produced through, 1) the decay of $\pi^0$ produced from interactions between CR nuclei and interstellar medium (ISM), 2) inverse Compton (IC) scattering of CR electrons/positrons with interstellar radiation field (ISRF), and 3) bremsstrahlung radiation of the electrons/positrons when interacting with ISM. Because they are not deflected by the magnetic field during propagation, these $\gamma$-rays can trace the distribution of CR sources, ISM, ISRF and so on. They are thus regarded as a very powerful probe to study the origin, propagation and interaction of CRs.

The propagation of CRs in the Galaxy is usually described by a diffusive process that charged particles are scattered by the random Galactic magnetic fields. Interactions, energy losses or regain are also considered in the propagation processes. A comprehensive description of the cosmic rays propagation and diffusive $\gamma$-ray emission via interaction between CRs and the ISM are developed as a computer package, GALPROP (Strong & Moskalenko 1998; Moskalenko & Strong 1998). Through fitting various kinds of observed CR data, one can determine the source parameters as well as the propagation parameters to fairly good precision. Strong et al. (2000) built a model which can basically fit the local CR data and EGRET all-sky $\gamma$-ray data except the well known "GeV excess" (Hunter et al. 1997). This model is referred as the "conventional" CR propagation model. The "GeV excess" problem has attracted lots of attention. Explanations range from the instrumental effect (Stecker et al. 2008), changes of CR flux spectra or normalizations (Gralewicz et al. 1997; Mori 1997; Porter & Protheroe 1997; Pohl & Esposito 1998; Aharonian & Atoyan 2000; Strong et al. 2000, 2004) to contribution of DM annihilation (de Boer 2005; de Boer et al. 2005; Bi et al. 2008a,b). However, new data from Fermi-LAT do not support the existence of "GeV excess" (Porter 2009; Abdo et al. 2009a). Another fascinating phenomenon is "Fermi Haze" in the inner Galaxy, which is assumed to be possibly the high energy counterpart to the microwave WMAP haze (Dobler et al. 2009). Whether "Fermi Haze" exists or not requires a more detailed model of diffuse $\gamma$-rays, because systematic effects in the analysis procedure of Dobler et al. (2009) could artificially induce such a similar "Haze" as pointed out by Linden & Profumo (2010).

Recently the observations on the CR positron fraction by PAMELA experiment (Adriani et al. 2009a) and the total electron and positron spectra by ATIC (Chang et al. 2008), PPB-BETS (Torii et al. 2008), H.E.S.S. (Aharonian et al. 2008; H. E. S. S. Collaboration: F. Aharonian 2009) and Fermi-LAT (Abdo et al. 2009b) all show interesting excesses when comparing with the conventional astrophysical background. Many possible origins have been proposed to explain the results, including astrophysical scenarios (e.g., Hooper et al. 2009; Yüksel et al. 2009; Profumo 2008; Hu et al. 2009; Fujita et al. 2009), and the models with exotic new physics like DM (e.g., Bergström et al. 2008; Barger et al. 2009; Cirelli et al. 2009; Yin et al. 2009). Since the electrons have important contribution to the diffuse $\gamma$-ray emission especially at high Galactic latitudes, it is necessary to revisit the problem of diffuse Galactic $\gamma$-ray emission after incorporating the new measurements of the electron spectrum. On the other hand, the diffuse $\gamma$-ray spectra can also be taken as a useful tool to probe the origin of the electron/positron excesses, which has been actually shown by some works (e.g., Bertone et al. 2009; Zhang et al. 2009; Bergström et al. 2009; Borriello et al. 2009). However, in these works, only limited sky regions such as the Galactic center are discussed. Furthermore in their discussion of DM models the contribution from substructures, which are definitely indicated by high resolution simulations (Diemand et al. 2005; Springel et al. 2008a), is not included. Therefore it would be valuable to revisit the diffuse $\gamma$-ray emission in a more detailed way.

In addition to the new electron measurements, new diffuse $\gamma$-ray emission results have been reported gradually by Fermi-LAT. Besides falsification in the "GeV excess", another result of importance is the featureless isotropic diffuse $\gamma$-ray emission (Abdo et al. 2010). This new measurements of the isotropic diffuse $\gamma$-ray component will be used in this work as the "extragalactic diffuse $\gamma$-ray background" (EGRB).

In this work, we investigate the diffuse $\gamma$ rays taking into account the three most popular kinds of models which can explain the electron/positron excesses: Galactic pulsars, DM annihilation and DM decay. These models represent three typical types of spatial distributions of the electron/positron sources, i.e., disc-like distribution (pulsars), highly concentrated spherical halo (DM annihilation) and less concentrated spherical halo (DM decay) respectively. In the case of DM annihilation, DM substructures from the recent most high resolution simulation (Springel et al. 2008a,b) are included in our study. EGRET data (Sreekumar 1995; Kniffen et al. 1996; Hunter et al. 1997) and Fermi-LAT latest observations around the Galactic center[1] (GC) (Digel 2009) and at mid-to-high latitude (Abdo et al. 2010) are used to compare with theoretical predictions.

In the following, we will first describe the conventional propagation model, which is used as the background, and its prediction of diffuse $\gamma$-rays in Sec. 2. Then we introduce the three kinds of models to reproduce the CR electron/positron excesses and present the corresponding all-sky diffuse $\gamma$-ray spectra in Sec. 3. The constraints on the models are discussed. Sec. 4 briefly introduces another adjusted background model which can reproduce

---

[1]https://confluence.slac.stanford.edu/display/LSP/Fermi+Symposium+2009

Fermi-LAT diffuse $\gamma$-ray data. Finally we give the conclusion and discussion in Sec. 5.

## 2. Cosmic rays and diffuse gamma rays in the conventional propagation model

The conventional CR propagation model predicts proton and electron spectra consistent with the local observations of CRs[2]. In this work we adopt GALPROP to calculate the propagation of CRs and the production of Galactic diffuse $\gamma$-rays. The propagation parameters in the conventional model are adjusted to reproduce CR data such as B/C, $^{10}$Be/$^9$Be, the local proton and electron spectra. The half height of the propagation halo is $z_h = 4$ kpc, which is the same as that taken in Strong et al. (2004). The diffusion + reacceleration model in Yin et al. (2009) is adopted in this work. The diffusion coefficient $D_{xx} = \beta D_0 (\rho/\rho_0)^\delta$ with $D_0 = 5.5 \times 10^{28}\,\mathrm{cm^2\,s^{-1}}$, $\delta = 0.34$. The Alfven speed is $v_A = 32$ km s$^{-1}$. The injection spectra of nuclei share the same power law form in rigidity with power index 1.94/2.42 below/above the break rigidity 15 GV and nuclei up to $Z = 28$ with all relevant isotopes are included. The injection spectra of primary electrons are 1.50/2.54 below/above the rigidity of 4 GV. Note that in order to be better consistent with the newly observed electron/positron data after including the extra component (see below and also Zhang et al. 2009), we adopt a slightly lower normalization ($\sim 0.9$) of the primary electrons compared with the conventional model given in Strong et al. (2004). In Fig. 1 we show the model predictions of B/C ratio, $^{10}$Be/$^9$Be ratio, local proton and electron spectra compared with data.

To describe the diffuse $\gamma$-ray emission at all directions we follow Strong et al. (2004) to divide the whole sky into six regions: region A (with Galactic longitude and latitude range of $330° < l < 30°$ and $0° < |b| < 5°$) corresponds to the "inner Galaxy", region B ($30° < l < 330°$ and $0° < |b| < 5°$) is the Galactic plane excluding the inner Galaxy, region C ($90° < l < 270°$ and $0° < |b| < 10°$) is the "outer Galaxy", regions D ($0° < l < 360°$ and $10° < |b| < 20°$) and E ($0° < l < 360°$ and $20° < |b| < 60°$) cover middle and high latitudes at all longitudes respectively, and region F ($0° < l < 360°$ and $60° < |b| < 90°$) describes the "Galactic poles". The calculated diffuse $\gamma$-ray spectra are shown in Fig. 2. We define the diffuse $\gamma$-ray emission predicted in this conventional model as the "diffuse $\gamma$-ray background", relative to the possible "diffuse $\gamma$-ray signals" from the extra electron/positron sources. The diffuse $\gamma$-ray background includes the contributions from decay of $\pi^0$ produced by nuclei collisions, IC scattering off the interstellar radiation field (ISRF), bremsstrahlung by electrons, and the isotropic EGRB. In this work, the EGRB from 30 MeV to 100 GeV is adopted as the fitted power law form $9.6 \times 10^{-3} E^{-2.41}$ cm$^{-2}$ s$^{-1}$ sr$^{-1}$ MeV$^{-1}$ according to the isotropic diffuse $\gamma$-ray spectrum given in Abdo et al. (2010). The EGRB becomes more and more predominant as increasing of the Galactic latitude. The diffuse $\gamma$-ray emission of this conventional model at all directions are roughly consistent with the EGRET data (except the "GeV excess"). However it somewhat underestimates the Fermi-LAT data at higher energies[3]. In the next section we will investigate the contribution from the extra electrons/positrons. We will show that this underestimate might be partially compensated by the contribution from the new electron/positrons sources.

---

[2]In this paper, if not specially stated, the CR data refers to the old observations before the new electron/positron measurements of PAMELA, PPB-PETS, ATIC, FERMI, H.E.S.S..

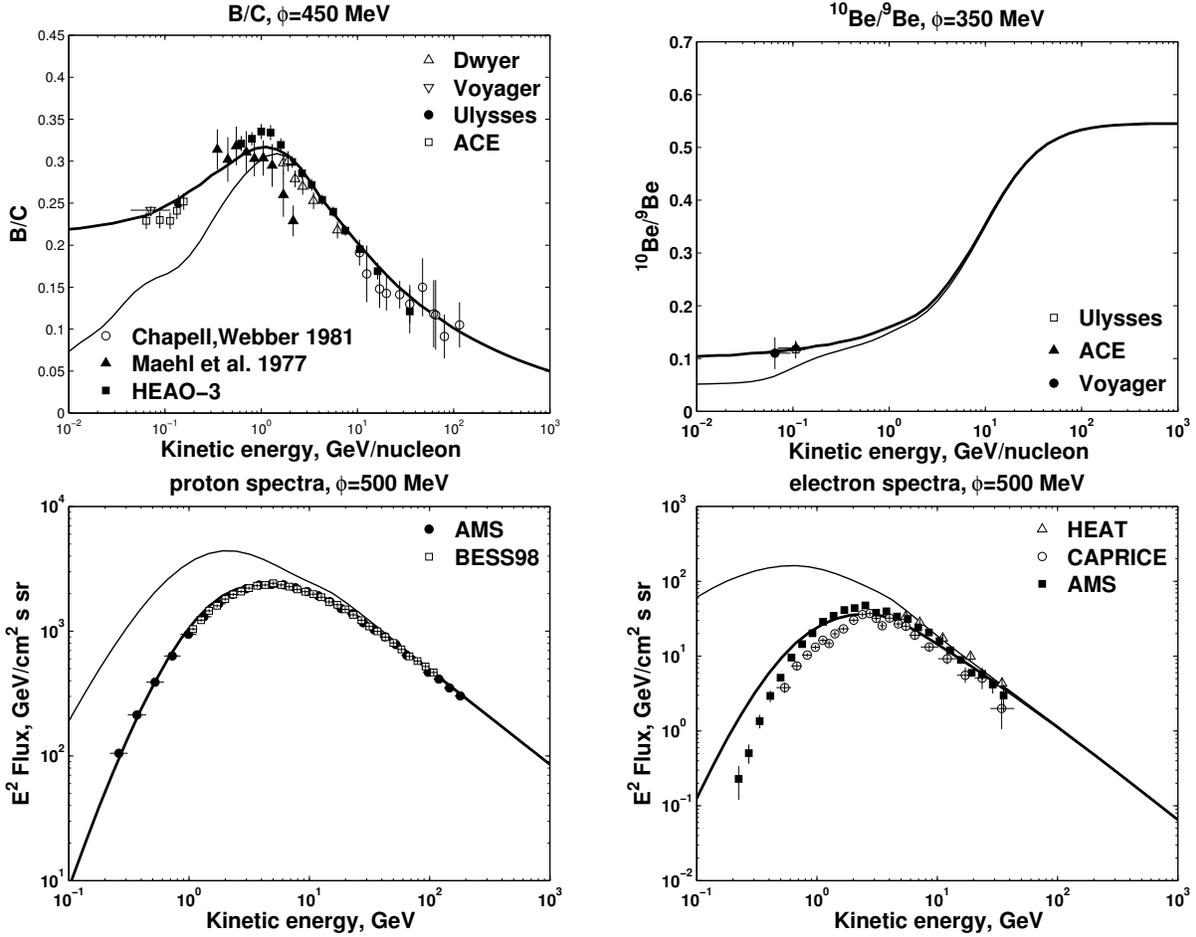

Fig. 1.— CR results in the conventional model. For B/C and $^{10}$Be/$^9$Be ratios the lower and upper curves correspond to the local interstellar (LIS) result and the solar modulated one respectively. On the contrary, for proton and electron spectra the lower curves are the modulated ones and the upper curves are for LIS. B/C data are from Chappell & Webber (1981), Maehl et al. (1977), Voyager (Lukasiak 1999), HEAO-3 (Engelmann et al. 1990), Ulysses (Duvernois et al. 1996), Dwyer (1978), ACE (Davis et al. 2000); $^{10}$Be/$^9$Be data from Ulysses (Connell 1998), ACE (Binns 1999) and Voyager (Lukasiak 1999); proton data from AMS (Alcaraz et al. 2000) and BESS98 (Sanuki et al. 2000); electron data from HEAT (Barwick et al. 1998) and CAPRICE (Boezio et al. 2000).

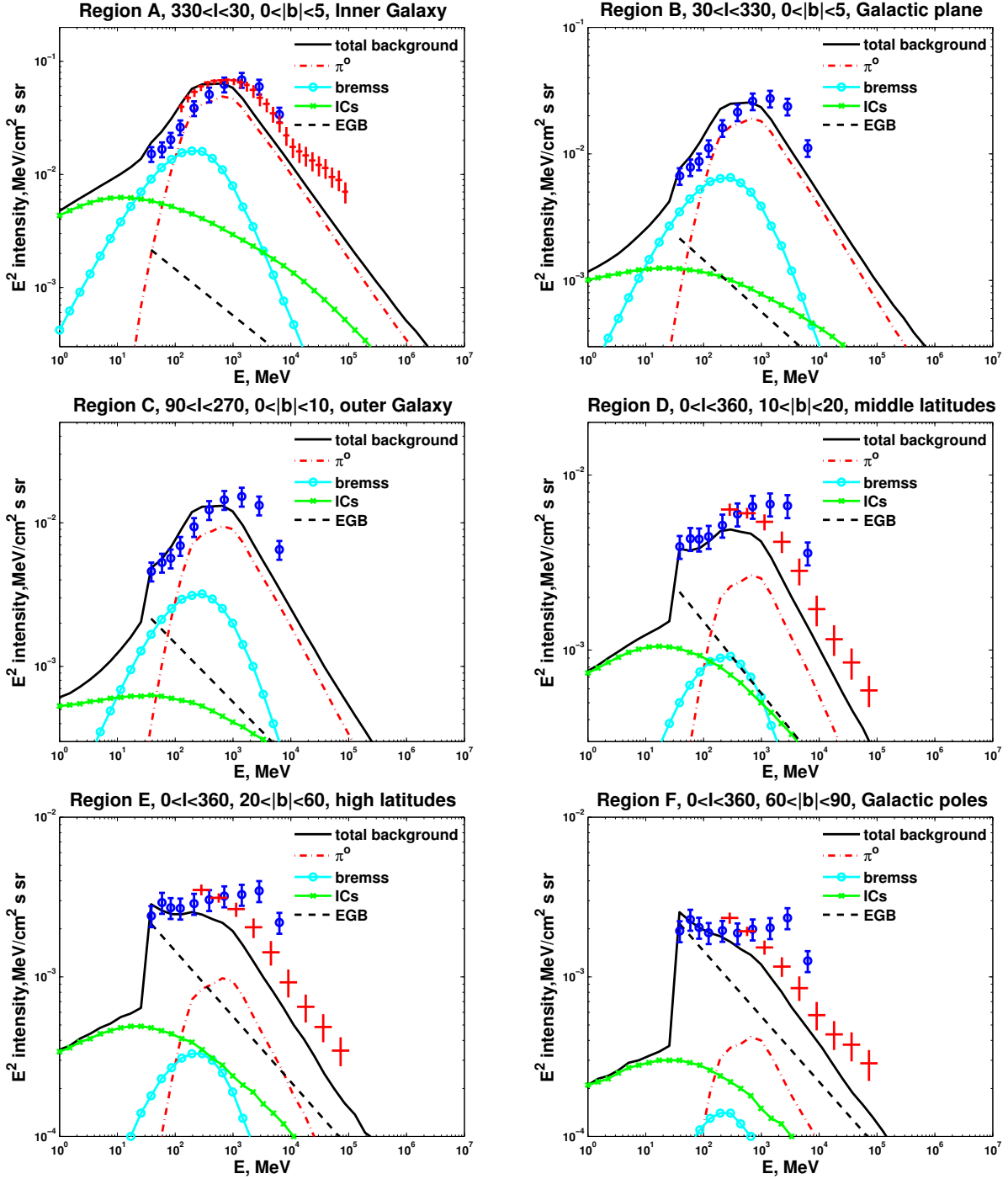

Fig. 2.— Spectra of diffuse γ-rays predicted by conventional propagation model for different sky regions. Data are from EGRET (blue circle, Sreekumar 1995; Kniffen et al. 1996; Hunter et al. 1997), and Fermi-LAT (red cross, Digel 2009; Abdo et al. 2010).

## 3. Scenarios accounting for the electron/positron excesses and the corresponding contributions to diffuse $\gamma$-rays

We refer to the conventional CR propagation model mentioned above as the "background model". The predicted electron flux in this background model is referred as the "electron background", and the diffuse $\gamma$-rays as the "diffuse $\gamma$-ray background". As we have shown in Sec. 2 the electron background can basically reproduce the old local observations. However, the recent experimental results of the positron fraction (Adriani et al. 2009a), and electron + positron spectra from ATIC (Chang et al. 2008), PPB-BETS (Torii et al. 2008), Fermi-LAT (Abdo et al. 2009b) and H.E.S.S. (Aharonian et al. 2008; H. E. S. S. Collaboration: F. Aharonian 2009) all show clear excesses above the predictions of the conventional model. Although there is freedom of the background contribution (Delahaye et al. 2009; Grasso et al. 2009), it is shown to be very difficult to explain the positron fraction and electron + positron spectra data simultaneously just using the background contribution (Grasso et al. 2009).

A direct way to solve the anomalies is to add new primary sources which can generate electron-positron pairs. It is found that, by adopting proper spectrum of the primary electrons and positrons and the conventional electron background, the ATIC or Fermi spectrum and the PAMELA positron ratio can be well reproduced simultaneously (e.g., our previous work Zhang et al. 2009). One should also note that the $\bar{p}/p$ ratio predicted by the conventional model is well consistent with the PAMELA data (Adriani et al. 2009b), which means the extra primary electron/positron sources should be leptonic. The electrons/positrons from the extra sources also induce diffuse $\gamma$-rays when propagating in the Galaxy. The scenarios usually used to explain the electron/positron excess will be described below as well as the corresponding contribution to diffuse $\gamma$-rays. The propagation of these primary electrons/positrons from the extra sources will also be solved in the conventional model.

---

[3]Note that to build a precise background model one may need the precise all-sky data from Fermi-LAT. Since the current available data are preliminary and there might be point sources contamination of the diffuse emission, we adopt this rough background model in this study.

### 3.1. Galactic pulsars

The electron/positron excesses observed at the CR experiments may come from some nearby astrophysical sources. A possible candidate of such sources is the Galactic pulsars. Pulsars and their nebulae are well known cosmic particle accelerators. Although the quantitative details of the acceleration processes are still open for study, early radio observations have established them as important high energy cosmic electron and positron sources (e.g., Frail et al. 1997). X-ray and $\gamma$-ray observations show that some of the accelerated particles can reach an energy of a few tens of TeV, and there are indications that the particle distribution cuts off in the TeV energy range (e.g., Pavlov et al. 2001; Lu et al. 2002; Aharonian et al. 2006). These particles produce emission over a broadband frequency range in the source region, which has been observed and studied extensively (e.g., Li et al. 2008). Some of these particles will escape into the ISM becoming high energy cosmic electrons and positrons. It is shown that one or several nearby pulsars will be able to contribute enough positrons to reproduce the PAMELA data (Hooper et al. 2009; Yüksel et al. 2009) as well as ATIC data (Profumo 2008).

In our calculation the source term for Galactic pulsars is parametrized as

$$q(R, z, E)|_P = K f(R, z) \left. \frac{\mathrm{d}N}{\mathrm{d}E} \right|_P, \tag{1}$$

where $K$ is the normalization factor representing the total luminosity of pulsars, $f(R, z)$ is the pulsar spatial distribution with $R$ the Galactocentric radius and $z$ the distance away from the Galactic Plane, and $\frac{\mathrm{d}N}{\mathrm{d}E}|_P$ is the average electron/positron energy spectrum generated from pulsars. The spatial distribution of pulsars can be parametrized as

$$f(R, z) \propto \left( \frac{R}{R_\odot} \right)^a \exp\left[ -\frac{b(R - R_\odot)}{R_\odot} \right] \exp\left( -\frac{|z|}{z_s} \right), \tag{2}$$

in which the parameters are adopted as $a = 1.0$, $b = 1.8$ (Zhang & Cheng 2001). Different from the spherically symmetric form of DM distribution, pulsars are mainly concentrated at the Galactic Plane with $z_s \sim 0.2$ kpc. The primary electron/positron spectrum injected by pulsars is generally assumed to be a power law form with an exponential cutoff at high energies

$$\left. \frac{\mathrm{d}N}{\mathrm{d}E} \right|_P = E^{-\alpha} \exp\left( -\frac{E}{E_{\mathrm{cut}}} \right). \tag{3}$$

For the pulsar model, we only need to consider the ICs contribution to the diffuse $\gamma$-rays from the primary electrons/positrons. Since the pulsars are concentrated in the Galactic plane within the propagation halo, we simply adopt GALPROP to calculate the ICs contribution to diffuse $\gamma$-rays. To some extent the pulsar scenario is equivalent to a propagation model simply adopting a harder electron injection spectrum, because pulsars and the CR sources have similar spatial distributions.

## 3.2. Dark matter models

Observations show that the mass of the Galaxy is dominated by DM. Many DM candidates have been proposed in the literature (Jungman et al. 1996; Bertone et al. 2005). The most attractive DM candidate is the weakly interacting massive particles (WIMP), which exist naturally in these models beyond the standard model, such as neutralinos in the supersymmetric model or the KK particles in the universal extra dimension model (Jungman et al. 1996; Bertone et al. 2005). These DM particles may decay at a very slow rate or annihilate into standard model particles, such as $\gamma$-rays, positrons/electrons and so on. Therefore the positron and electron excesses at the CR experiments have been widely considered as possible signals from DM decay or annihilation (Bergström et al. 2008; Barger et al. 2009; Cirelli et al. 2009; Yin et al. 2009).

Similar to the case of Galactic pulsars we incorporate these electron/positron sources from DM annihilation or decay in the GALPROP package so that we can calculate the fluxes on the Earth. We use the same conventional propagation model as described in the previous section to calculate the contribution to local electron/positron observations from the DM sources. The source function for DM annihilation and decay can be written as

$$q(\mathbf{r}, E)|_A = \frac{\langle \sigma v \rangle}{2m_\chi^2} \left. \frac{\mathrm{d}N}{\mathrm{d}E} \right|_A \rho^2(\mathbf{r}), \tag{4}$$

$$q(\mathbf{r}, E)|_D = \frac{1}{m_\chi \tau} \left. \frac{\mathrm{d}N}{\mathrm{d}E} \right|_D \rho(\mathbf{r}), \tag{5}$$

where $m_\chi$ is the mass of DM particle, $\langle \sigma v \rangle$ (or $\tau$) is the annihilation cross section (or lifetime of DM particles), $\frac{\mathrm{d}N}{\mathrm{d}E}$ is the electron/positron yield spectrum for specified channel per annihilation or decay. In this work we assume the DM annihilation or decay directly into leptons, i.e., $\chi\chi \to e^+e^-, \mu^+\mu^-, \tau^+\tau^-$. PYTHIA package (Sjöstrand et al. 2006) is used to simulate the radiative and decaying processes of the final state particles. Finally $\rho(\mathbf{r})$ is the density profile of the Milky Way halo, for which we take an Einasto form (Einasto 1965)

$$\rho(r) = \rho_0 \exp\left[-\frac{2}{\alpha}\left(\frac{r^\alpha - R_\odot^\alpha}{r_{-2}^\alpha}\right)\right], \tag{6}$$

with $\alpha = 0.2$, $r_{-2} = 25$ kpc and the local DM density $\rho_\odot = 0.3$ GeV cm$^{-3}$ at $r = R_\odot \equiv$ 8.5 kpc. The total mass of the halo inside the virial radius $r_{\rm vir} \approx 220$ kpc, within which the mean density is 200 times the critical density, is measured to be $\sim 1.25 \times 10^{12}$ M$_\odot$. This profile is also able to fit the WMAP haze data (Finkbeiner 2004) assuming DM annihilation to produce electrons/positrons (Zhang et al. 2009).

We also include the DM substructures in our calculation according to the newest *Aquarius* N-body simulation of DM structure formation (Springel et al. 2008b). The annihilation luminosity, defined as $L \propto \int \rho^2 {\rm d}V$, of the substructures can be scaled with respect to the smooth halo as (Springel et al. 2008b; Pinzke et al. 2009)

$$L_{\rm sub}(< r) = 0.8 L_{\rm sm} \left( \frac{M_{\rm min}}{10^5\,{\rm M}_\odot} \right)^{-0.226} \left( \frac{r}{r_{\rm vir}} \right)^{0.8(r/r_{\rm vir})^{-0.315}}, \tag{7}$$

where $L_{\rm sm}$ is the luminosity of the smooth halo, $M_{\rm min}$ is the minimum subhalo mass corresponding to the free streaming of DM particles. As the canonical value for cold DM particles (Hofmann et al. 2001; Chen et al. 2001), $M_{\rm min} \approx 10^{-6}$ M$_\odot$. In this work, the most conservative case is taken, where only resolved subhalos in the simulation are considered, i.e. $M_{\rm min} \approx 10^5$ M$_\odot$. The mass fraction of the subhalos is estimated to be $\sim 18\%$ in *Aquarius* simulation (Springel et al. 2008a). Thus for the annihilation DM case, we rescale Eq. (6) with a factor of 0.82 when calculating the contribution from the smooth halo. For the decaying DM case, there will be almost no effect for substructures.

The $\gamma$-rays produced in the DM models consist of two parts: the IC radiation of the electrons/positrons when scattering with ISRF and the cosmic microwave background (CMB), and the final state radiation (FSR) emitted directly from the external legs when DM particles annihilates or decays to charged leptons. For the $\tau^+\tau^-$ channel, the decay of tauons will produce a large number of neutral pions which can also decay into photons. This contribution is also included in the FSR.

The IC radiation is further divided into two components depending on whether the scatterings occur inside or outside of the CR propagation halo. For the electrons/positrons inside the propagation halo ($R < 15 kpc$, $z_h < 4 kpc$), the ISRF is composed of the star light in optical band, the dust radiation in infrared band and the CMB. We incorporate this component in GALPROP to calculate the diffuse $\gamma$-ray emission (labeled as $\phi_{\rm IC}^{\rm in}$). For the electrons/positrons out of the propagation halo only the CMB is relevant and the calculation is done separately. The energy loss dominates the diffusion process in the dark halo, therefore the equivalent spectrum of electrons/positrons out of the propagation halo can be written as (Colafrancesco et al. 2006)

$$\frac{{\rm d}n}{{\rm d}E} = \frac{1}{b(E)} \int_E^\infty {\rm d}E' q(\mathbf{r}, E'), \tag{8}$$

where $q(\mathbf{r}, E)$ is the source function given in Eqs.(4) and (5), $b(E) = 2.5 \times 10^{-17}(E/\text{GeV})^2$ GeV s$^{-1}$ is the IC energy loss rate of the electrons/positrons when scattering off the CMB photons. Then the IC photon emissivity is given by (Blumenthal & Gould 1970)

$$q_{\text{IC}}^{\text{out}}(\mathbf{r}, E_\gamma) = \int \text{d}\epsilon \, n(\epsilon) \int \text{d}E \frac{\text{d}n}{\text{d}E} \times F(\epsilon, E, E_\gamma), \tag{9}$$

where $n(\epsilon)$ is the differential number density of CMB photon as a function of energy $\epsilon$. Function $F(\epsilon, E, E_\gamma)$ is given by

$$F(\epsilon, E, E_\gamma) = \frac{3\sigma_T}{4\gamma^2\epsilon} \left[ 2q \ln q + (1 + 2q)(1 - q) + \frac{(\Gamma q)^2(1 - q)}{2(1 + \Gamma q)} \right], \tag{10}$$

with $\sigma_T$ the Thomson cross section, $\gamma$ the Lorentz factor of electron, $\Gamma = 4\epsilon\gamma/m_e$, and $q = \frac{E_\gamma/E}{\Gamma(1 - E_\gamma/E)}$. For $q < 1/4\gamma^2$ or $q > 1$ we set $F(\epsilon, E, E_\gamma) = 0$. The flux is given through a line-of-sight integral over the emissivity

$$\phi_{\text{IC}}^{\text{out}} = \frac{1}{4\pi} \int q_{\text{IC}}^{\text{out}} \text{d}l. \tag{11}$$

The FSR is radiated in the whole DM halo. The photon yield spectrum for $e^+e^-$ or $\mu^+\mu^-$ channel for $m_\chi \gg m_e$, $m_\mu$ can be written as (Bergström et al. 2005)

$$\left. \frac{\text{d}N}{\text{d}x} \right|_i = \frac{\alpha}{\pi} \frac{1 + (1 - x)^2}{x} \log \left( \frac{s}{m_i^2}(1 - x) \right), \tag{12}$$

where $\alpha \approx 1/137$ is the fine structure constant, $i = e, \mu$. For DM annihilation we have $s = 4m_\chi^2$ and $x = E_\gamma/m_\chi$; while for DM decay $s = m_\chi^2$ and $x = 2E_\gamma/m_\chi$ (Essig et al. 2009). For the $\tau^+\tau^-$ channel, there is internal bremsstrahlung radiation as shown in Eq.(12) as well as the decay products from the chain $\tau \rightarrow \pi^0 \rightarrow \gamma$. We adopt a total parametrization as (Fornengo et al. 2004)

$$\left. \frac{\text{d}N}{\text{d}x} \right|_\tau = x^{-1.31} \left( 6.94x - 4.93x^2 - 0.51x^3 \right) e^{-4.53x}, \tag{13}$$

with the same definition of $x$ as above.

The $\gamma$-ray flux along a specific direction can be written as

$$
\begin{aligned}
\phi_{\text{FSR}}(E_\gamma, \psi) &= C \times W(E_\gamma) \times J(\psi) \\
&= \begin{cases} \frac{\rho_\odot^2 R_\odot}{4\pi} \times \frac{\langle \sigma v \rangle}{2m_\chi^2} \frac{\text{d}N}{\text{d}E_\gamma} \times \frac{1}{\rho_\odot^2 R_\odot} \int_{LOS} \rho^2(l) \text{d}l, & \text{for annihilation} \\ \frac{\rho_\odot R_\odot}{4\pi} \times \frac{1}{m_\chi \tau} \frac{\text{d}N}{\text{d}E_\gamma} \times \frac{1}{\rho_\odot R_\odot} \int_{LOS} \rho(l) \text{d}l, & \text{for decay} \end{cases}
\end{aligned} \tag{14}
$$

where the integral is taken along the line-of-sight, $W(E)$ and $J(\psi)$ represent the particle physics factor and the astrophysical factor respectively, and $\frac{\text{d}N}{\text{d}E_\gamma} = \frac{1}{m_\chi} \frac{\text{d}N}{\text{d}x}$.

### 3.3. Results

In this subsection, we will present the results of diffuse $\gamma$-ray spectra after taking into account the extra electron/positron sources required by the new CR electron data. Different scenarios are firstly normalized to fit the local PAMELA and ATIC or Fermi-LAT observations based on the conventional propagation model described in Section 2. Then the $\gamma$-ray emissivity by the primary sources are calculated. The emission at a direction is calculated by integrating the emissivity of $\gamma$-rays along the line of sight.

As there are discrepancies for the electron + positron spectra between ATIC and Fermi-LAT, we fit the two data sets without any bias. The first one is the combination of PAMELA and ATIC data, and the second one is PAMELA + Fermi-LAT + H.E.S.S. data. The model parameters of the primary sources are accordingly different to fit these two data sets.

For PAMELA + **A**TIC data, the model parameters are:

- **Aanni**: for DM **anni**hilation, $m_\chi = 1$ TeV, annihilation cross section $\langle\sigma v\rangle = 3.6\times10^{-23}$ cm$^3$ s$^{-1}$ corresponding to a boost factor $\sim 1200$ with respect to the natural value $\langle\sigma v\rangle_0 = 3\times10^{-26}$ cm$^3$ s$^{-1}$, and the annihilation channels are $e^\pm$, $\mu^\pm$ and $\tau^\pm$ with equal branching ratios.

- **Adecay**: for DM **decay**, $m_\chi = 2$ TeV with life time $\tau \sim 10^{26}$ s, and also equal branching ratios into $e^\pm$, $\mu^\pm$ and $\tau^\pm$.

- **Apsr**: for **puls**a**r**s, we adopt $\alpha = 1.0$ and $E_{\rm cut} \sim 600$ GeV.

While for PAMELA + **F**ermi-LAT + H.E.S.S. data, the electron/positron spectra observed by Fermi-LAT and H.E.S.S. are softer and smoother than that from ATIC, which favor a softer contribution from the extra component. For the DM models we will consider pure muon or tauon final states respectively. We adopt the following model parameters:

- **Fanni**: for DM **anni**hilation, $m_\chi = 1.7$ TeV, cross section $\langle\sigma v\rangle = 5.4 \times 10^{-23}$ cm$^3$ s$^{-1}$ for pure $\mu^\pm$ channel, and $m_\chi = 3$ TeV, $\langle\sigma v\rangle = 19.0 \times 10^{-23}$ cm$^3$ s$^{-1}$ for pure $\tau^\pm$ channel.

- **Fdecay**: for DM **decay**, $m_\chi = 3.4$ TeV with life time $\tau \sim 1.45 \times 10^{26}$ s for pure $\mu^\pm$ channel and $m_\chi = 6$ TeV with life time $\tau \sim 0.61 \times 10^{26}$ s for pure $\tau^\pm$ channel.

- **Fpsr**: for **puls**a**r**s, $\alpha = 1.4$ and $E_{\rm cut} \sim 800$ GeV.

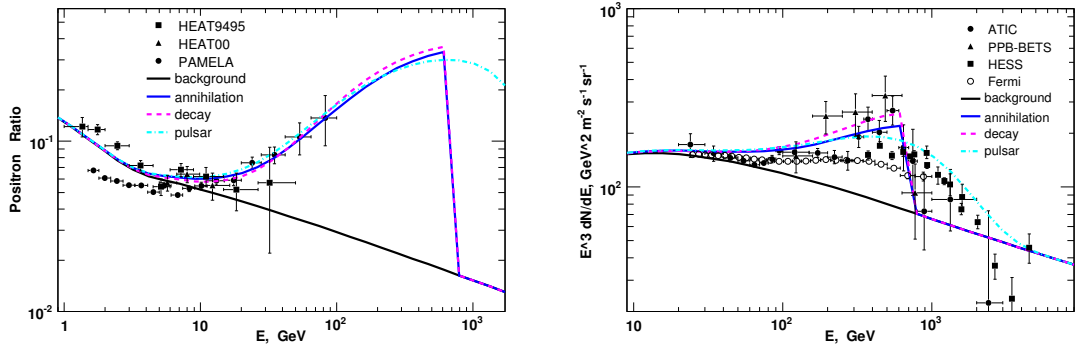

Fig. 3.— *Left:* The positron fractions predicted by the three scenarios after solar modulation, compared with HEAT (Barwick et al. 1997; Coutu et al. 2001) and PAMELA (Adriani et al. 2009a) data. *Right:* The total electron + positron fluxes of the three scenarios, compared with observations of ATIC (Chang et al. 2008), PPB-BETS Torii et al. (2008), H.E.S.S. (Aharonian et al. 2008; H. E. S. S. Collaboration: F. Aharonian 2009) and Fermi-LAT (Abdo et al. 2009b).

Fig. 3 illustrates how the PAMELA and ATIC observations are explained by adding extra primary sources, the pulsars, DM annihilation and decay, respectively. Using the second sets of parameters above, PAMELA+Fermi-LAT and H.E.S.S. electrons could also be well reproduced. For clarity, only plotted are parameters fitted to PAMELA and ATIC observations. In summary, by taking proper parameters for these different primary sources the local measurements of the electron/positron spectrum can be well explained. However, as the spatial distributions of these primary sources are quite different, we expect the corresponding diffuse $\gamma$-ray spectra should be not alike.

The full-sky diffuse $\gamma$-ray spectra of these models are given in Figs. 4 and 5. In Fig. 4, DM models used to fit the PAMELA + Fermi-LAT + H.E.S.S. data is with pure $\mu^{\pm}$ channel, while in Fig. 5 these DM models is with pure $\tau^{\pm}$ channel. To better illustrate the contributions of each component, we plot the detailed results in Figs. 6 and 7 for regions A and D respectively, taking the $\mu^{\pm}$ and $\tau^{\pm}$ final states for **Fanni** model as examples.

From these figures, we notice that the diffuse $\gamma$-ray contribution from pulsar scenario is tiny and almost negligible. While for DM scenarios, there is always a bump at energies from several hundred GeV to TeV due to the FSR. This spectral feature is more remarkable in high latitude sky regions. And this high energy bump is extremely clear for DM models with tauon final states as can be seen in Fig. 5. Further more, the IC diffuse $\gamma$-ray component of DM models is relatively more significant in the high latitude regions (e.g., E and F) where the Galactic background is lower. They can somewhat compensate the underestimation of diffuse $\gamma$-rays by background propagation model relative to the Fermi-LAT observations. These IC components mainly contribute at the energy range less than several hundred GeV. The IC component of DM models inside the diffusion halo, resulting from interactions with ISRF, contributes at higher energies than those outside the diffusion halo which results from interactions with CMB. Because we take the most conservative minimum subhalo mass, the latter IC contribution is not large enough to be shown in Figs. 6 and 7.

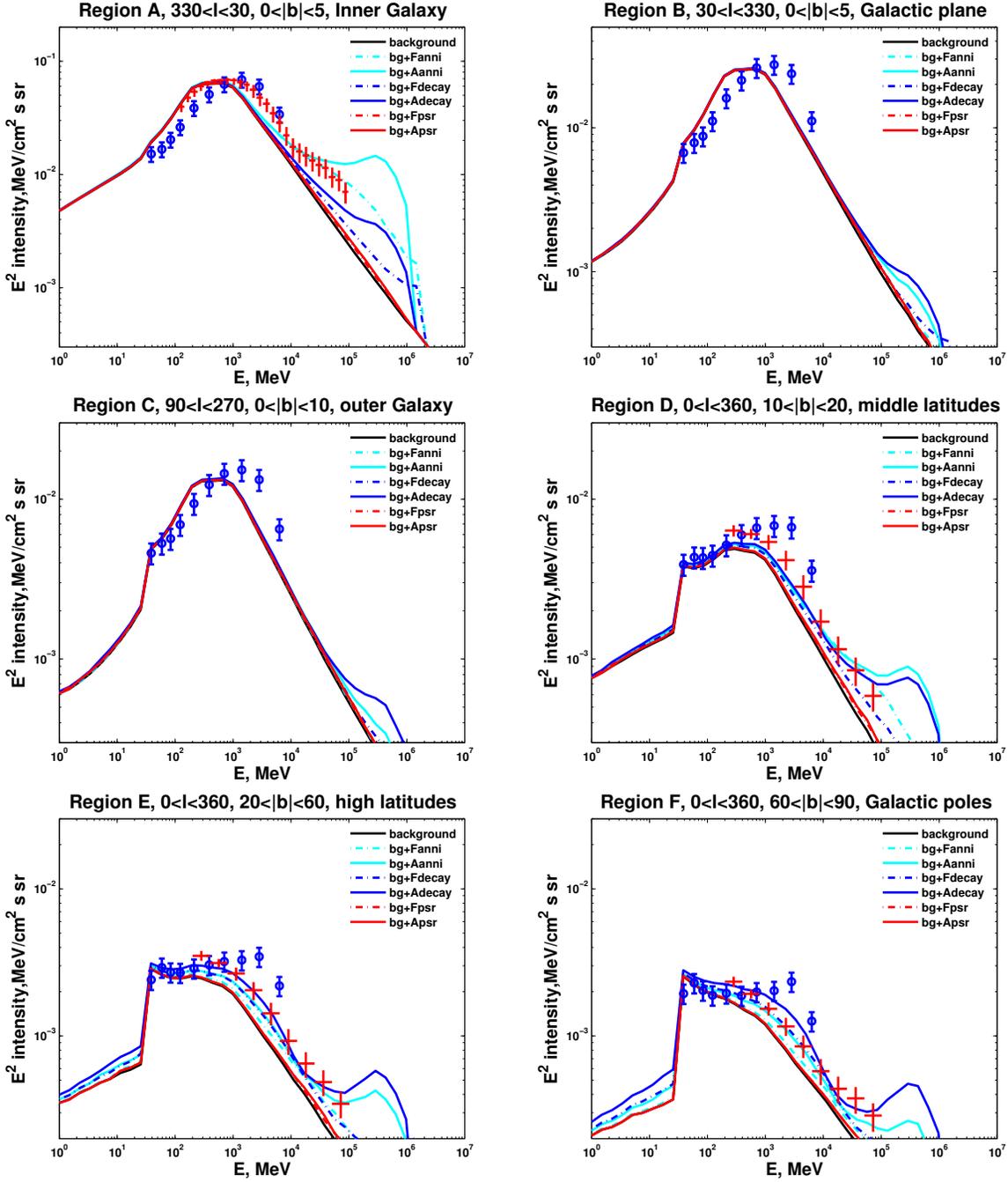

Fig. 4.— The full-sky diffuse γ-ray spectra. In this plot, the final state of **Fanni** and **Fdecay** DM models is pure $\mu^{\pm}$. The black-solid line denotes the background. The observation data are the same as in Fig. 2.

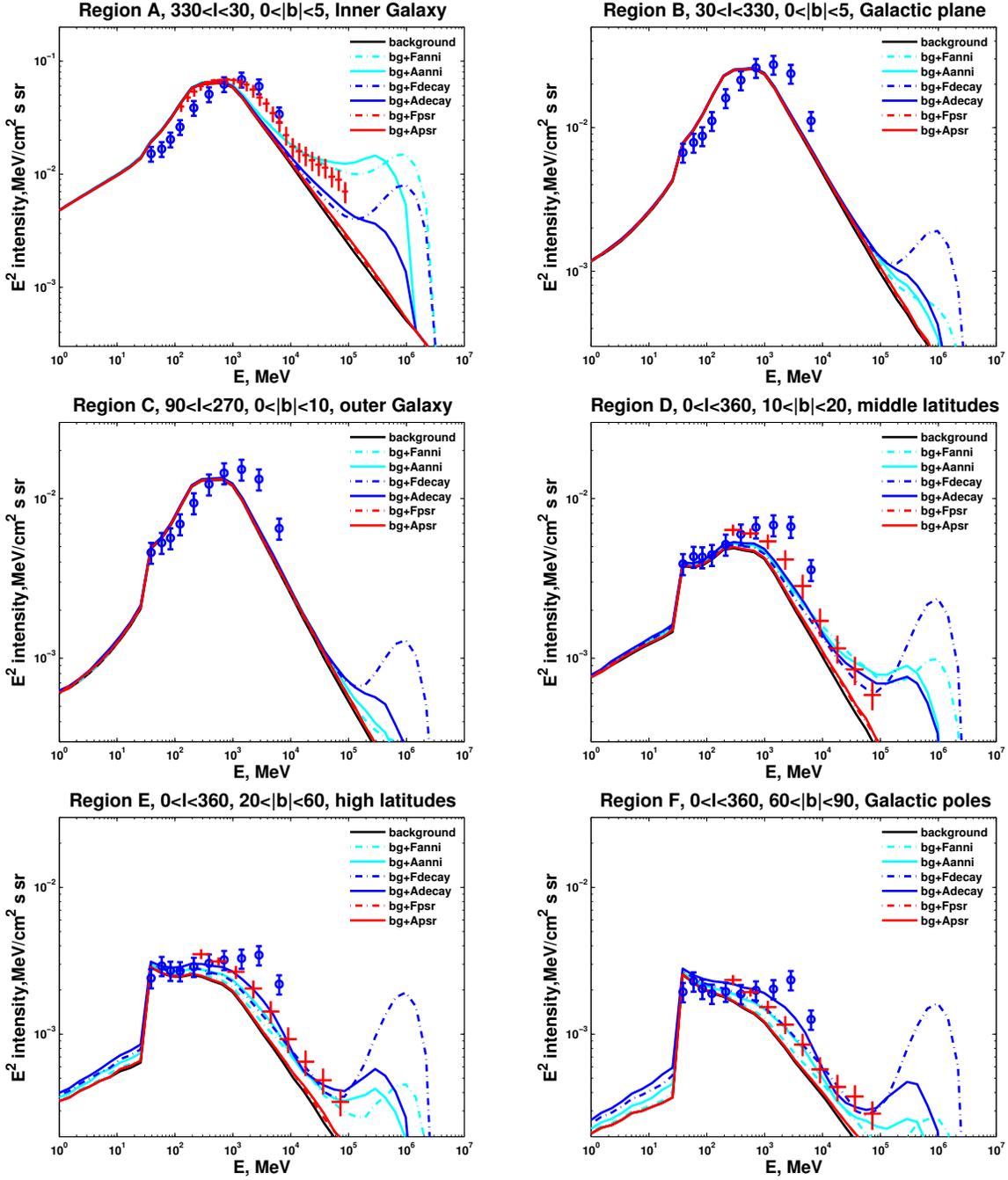

Fig. 5.— Same as Fig. 4 except that the final state of **Fanni** and **Fdecay** DM models is pure $\tau^{\pm}$.

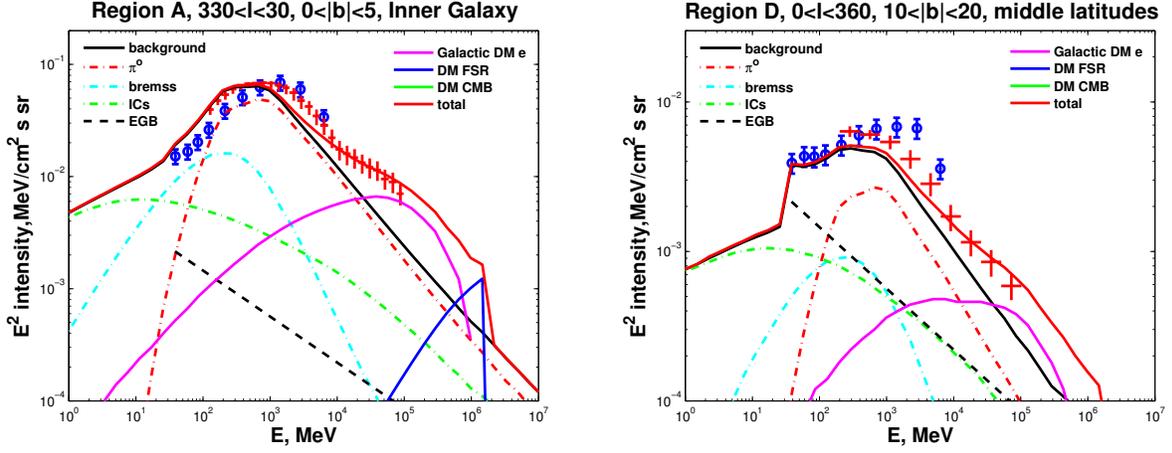

Fig. 6.— Contributions of each component for **Fanni** model with $\mu^{\pm}$ final state. Black solid line: total background with $\pi^0$ component (red dot-dashed), IC (green dot-dashed), bremsstrahlung (cyan dot-dashed) and the isotropic extragalactic background (black dashed); magenta solid line: IC of DM electrons inside the diffusion halo; blue solid line: FSR from DM; green solid line: IC of DM electrons outside the diffusion halo. The sum of these components are represented in red solid line.

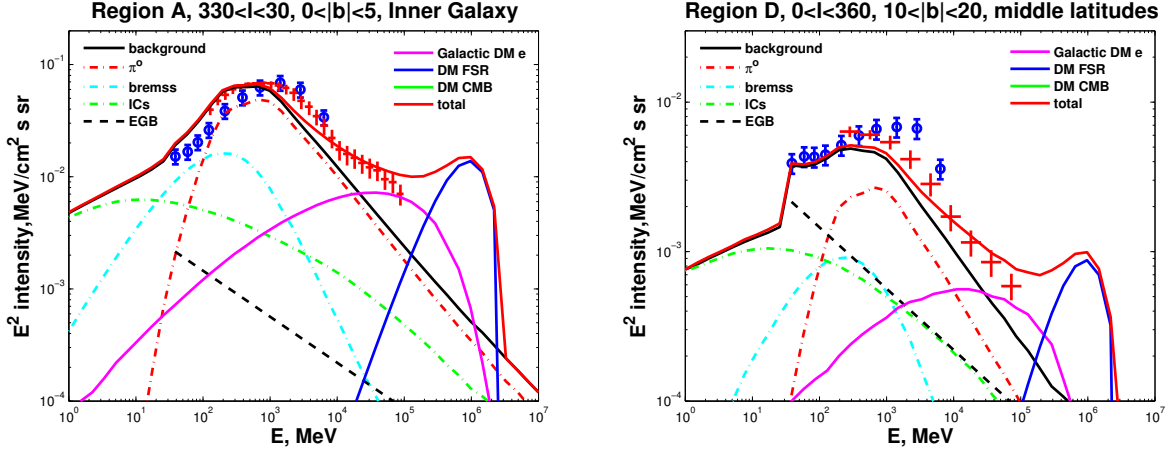

Fig. 7.— Same as Fig. 6, but for $\tau^{\pm}$ final state of DM annihilation.

Current observations can already constrain or exclude some of the model configurations. In the inner Galaxy (Region A), Figs. 4 and 5 show that predictions of **Aanni** and **Fanni** with pure $\tau^{\pm}$ channel have already exceeded the high energy Fermi-LAT data. The **Fanni** model with pure $\mu^{\pm}$ channel can survive the constraint. Comparing Fig. 6 with 7, we know that the tension is ascribed to the inclusion of $\tau^{\pm}$ in the final states. DM annihilation models with $\tau^{\pm}$ as one of the main final states would produce too many diffuse $\gamma$-rays, thus disfavored by current measurements.

The constraint on the decaying DM model is much weaker. However, we can see that the data from the Galactic pole (Region F) can more or less constrain **Adecay** model. But the exclusion power for decaying DM scenario is rather weak. Combining the results of Region A, we find that **Fdecay** scenario or **Fanni** with $\mu^{\pm}$ final state can still safely survive the current Fermi-LAT data.

It is interesting to note that the underestimate of the background diffuse $\gamma$-rays can be filled to some extent by DM contribution such as **Fdecay** scenario. It is to say there is possibility that the excesses of electrons/positrons and Fermi-LAT diffuse $\gamma$-rays are of the common DM origin. However, a more definite conclusion needs better understanding of the background contribution. In the next section we will show that how an adjusted background can reproduce the Fermi-LAT diffuse $\gamma$-ray data.

## 4. Adjusted background model fitting Fermi-LAT diffuse $\gamma$-ray data

In the above sections, the background propagation model is chosen so that it can reproduce the CR data and, more importantly, can explain the new electron/positron observations with extra sources. But such a conventional model fails to match the all-sky Fermi-LAT diffuse $\gamma$-ray data. Considering the possible large fluctuations of electron spectrum in the Galactic diffusion region originating from its rapid energy losses, the stochastic sources, propagation processes and so on (Strong et al. 2004), it is unnecessary for electrons to be normalized to the local observations. We find that with the normalization of electron flux three times larger, based on another conventional model (Bi et al. 2008a), the background could fit the Fermi data as well, as shown in Fig. 8. The fit to mid-to-high latitude data are good, but in the GC direction, the fit is not so good above 1 GeV. The possible reason is that the GC is a complex region, and the current Fermi-LAT data still suffer from point source contamination, especially in the GC region. This adjusted model is given to just roughly illustrate that the Fermi-LAT diffuse $\gamma$-ray data might be reproduced by a background CR propagation model. In such a case, there would be much less space for DM signals. That is to say, the constraints on DM models will be more severe.

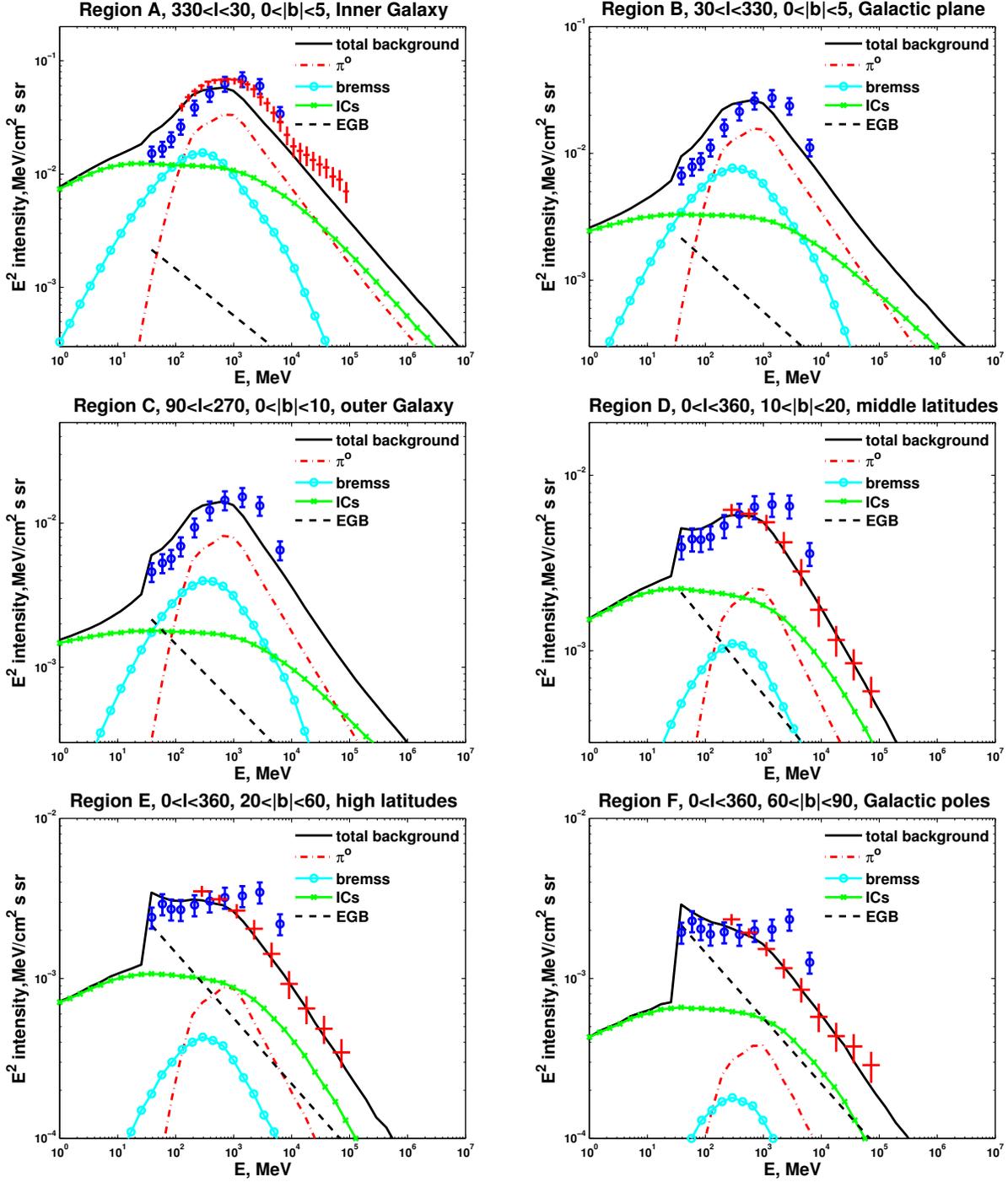

Fig. 8.— Spectra of diffuse γ-rays for different sky regions for an adjusted background model, which is based on the conventional one in Bi et al. (2008a) but with electron normalization three times larger. Data and lines are the same as in Fig. 2.

## 5. Discussion and Conclusion

In summary, we recalculate the all-sky diffuse $\gamma$-ray emission in this work taking the recent CR $e^{\pm}$ excesses into account. We consider three typical models: Galactic pulsars, DM annihilation and DM decay. For each model, we adopt typical parameter settings which can reproduce the new electron/positron observations. We then investigate the diffuse $\gamma$-ray emission, generated through IC scattering of the electrons/positrons off the ISRF and/or CMB, and the FSR from DM annihilation or decay. For the DM annihilation model, the substructure effect is included according to the newest high resolution N-body simulation.

Our results show that, the contribution to diffuse $\gamma$-rays from pulsars is always negligible in any region of the sky. While for DM models, the current Fermi-LAT data can set constraints on the models. Specifically, the inner Galaxy data can constrain annihilating DM models **Fanni** with $\tau^{\pm}$ final state and **Aanni** models. The constraint on the decaying DM model is much weaker. However, data from the Galactic pole can constrain **Adecay** to some extent. Furthermore there is a prominent high energy bump arising from FSR for both annihilating and decaying DM scenarios.

The IC component of DM models could somewhat compensate the underestimate of diffuse $\gamma$-rays from background propagation models. It contributes from sub-GeV to several hundred GeV range, and is relatively more significant at higher latitudes than at lower latitudes. This provides us a possibility that a proper DM model (e.g., **Fdecay**) can explain the electron/positron excesses and Fermi-LAT diffuse $\gamma$-rays simultaneously. However, it may be too early to reach such a conclusion due to the possible large uncertainty about the background model. We show that an adjusted propagation model could fit Fermi-LAT diffuse $\gamma$-ray data well, considering the large fluctuation of electron spectrum in the diffusion region. With more precise data and less contamination of sources, we may get closer to derive a precise background model. If a refined background CR propagation model could indeed explain all CR and diffuse $\gamma$-ray data, constraints on DM models would be stronger.

Finally we discuss some words about the DM subhalos. In this work, we conservatively use the resolved subhalos in the high resolution simulation by Springel et al. (2008b), i.e. $M_{\min} \sim 10^5$ M$_\odot$. It can be expected that there might be many unresolved subhalos in the Galactic halo. The extrapolation according to the statistical results of the resolved subhalos indicates that the unresolved ones play a much more significant role in the annihilation luminosity (Eq. (7)). That is to say if the unresolved subhalos are properly taken into account, the current constraint on annihilating DM model would be even stronger. On the other hand, diffuse $\gamma$-rays can also be used as a tool to explore how low the subhalo mass that we can extrapolate is. This might be a way to approach the DM particle nature from structure information (Bi et al. 2009; Yuan et al. 2010).

BXJ thanks I. V. Moskalenko for helpful discussion during the *1st workshop on gamma ray astronomy at high altitude* at Hebei, China. ZJ thanks Qiao Song, Yu Lu, Jun-qiang Ge and Ming Yang for their help with IDL and HEALPix. Specially, Zhang thanks Gregory Dobler and Douglas P. Finkbeiner sincerely for their kind help with Fermi data. This work is supported in part by the Natural Sciences Foundation of China (No. 10773011) and by the Chinese Academy of Sciences under the grant No. KJCX3-SYW-N2, and by the National Basic Research Program of China (973 Program) under grant No. 2010CB833000.